# Estimation of the Evolutionary Spectra With Application to Stationarity Test

Yu Xiang , *Member, IEEE*, Jie Ding , *Member, IEEE*, and Vahid Tarokh, *Fellow, IEEE*

*Abstract*—In this paper, we propose a new inference procedure for understanding non-stationary processes, under the framework of evolutionary spectra developed by Priestley. Among various frameworks of modeling non-stationary processes, the distinguishing feature of the evolutionary spectra is its focus on the physical meaning of frequency. The classical estimate of the evolutionary spectral density is based on a double-window technique consisting of a short-time Fourier transform and a smoothing. However, smoothing is known to suffer from the so-called bias leakage problem. By incorporating Thomson's multitaper method that was originally designed for stationary processes, we propose an improved estimate of the evolutionary spectral density, and analyze its bias/variance/resolution tradeoff. As an application of the new estimate, we further propose a non-parametric rank-based stationarity test, and provide various experimental studies.

## I. INTRODUCTION

**N**ONSTATIONARY processes are common across a variety of areas and serve as a natural generalization of the classical wide-sense stationary processes. Because of their wide range of applications, they have been an active research area in many different areas including signal processing, statistics, neuroscience, and economics.

However, the intrinsic complexity of the non-stationarity precludes a unique way of modeling the non-stationary processes. Various frameworks have been developed over the past few decades: instantaneous power spectra [1], evolutionary spectra [2], Wigner-Ville spectral analysis [3], locally stationary processes [4], and local cosine basis [5] among others. In this work, we adopt the *evolutionary spectra* framework developed by Priestley and his colleagues [2], [6]–[8], which is one of the first attempts to model non-stationary processes from the spectral point of view. The appealing aspect of this framework is its emphasis on the physical meaning of frequency, while generalizing the spectral representation of the stationary processes to that of the non-stationary processes [9].

Perhaps the most closely related framework is the *locally stationary processes* framework. Since Dahlhaus developed this framework in a series of papers [4], [10], [11], it has been extensively studied (see for example [12], [13] and references therein). We attempt to summarize the main differences between the two frameworks as follows. The evolutionary spectra framework is motivated by the physical interpretation of frequency, but does not guarantee the uniqueness of the spectral density. On the other hand, the locally stationary processes framework guarantees the uniqueness of the spectral density by providing an asymptotic analysis of the non-stationary processes. However, the rescaling technique, which is central to that framework, may sacrifice physical interpretations in some real applications. One related framework is developed based on SLEX (smooth localized complex exponential basis) functions. Interestingly, it is shown to be asymptotically mean square equivalent to Dahlhaus's framework [14]. These SLEX based methods are quite useful for very long time series, as the dyadic segmentation could help as a first approximation step. More comparisons between Priestley's and Dahlhaus's frameworks can also be found in [15] and detailed discussions on the other frameworks can be found in [9], [16] and the references therein.

The estimation procedure of the evolutionary spectra in [2] is based on the so-called double-window technique, consisting of a short-time Fourier transform and smoothing. However, the smoothing step is known to suffer from the so-called bias leakage problem [17]. To overcome this problem for stationary processes, various tapering methods have been developed and Thomson's *multitaper method* [18] is arguably the most widely used one. In this work, we apply the multitaper method to the estimation of evolutionary spectral density and analyze the bias/variance/resolution tradeoff of the estimate. We show that the non-stationarity calls for additional considerations of the tradeoff, which provides insights into window design, choice of frequency resolution and number of tapers. As an application of the estimate, we propose a non-parametric rank-based stationarity test and compare it with the stationarity test investigated by Priestley and Subba Rao in [7]. Our test is more robust to the underlying distribution of the data, and it can serve as a complementary test to the existing stationarity tests from our numerical experiments.

There are a few other related works in the literature. In [19], the authors expressed Priestley's two step approach in the form of the multitaper formulation, where the number of tapers

Manuscript received February 24, 2018; revised September 26, 2018 and October 29, 2018; accepted November 26, 2018. Date of publication January 1, 2019; date of current version January 16, 2019. The associate editor coordinating the review of this manuscript and approving it for publication was Prof. D. Robert Iskander. This work was supported by Defense Advanced Research Projects Agency (DARPA) under Grants W911NF-14-1-0508, W911NF-16-1-0561, and N66001-15-C-4028. *(Corresponding author: Yu Xiang.)*

Y. Xiang is with the Electrical and Computer Engineering Department, University of Utah, Salt Lake City, UT 84112 USA (e-mail: yu.xiang@utah.edu).

J. Ding is with the School of Statistics, University of Minnesota, Minneapolis, MN 55455 USA (e-mail: dingj@umn.edu).

V. Tarokh is with the Electrical and Computer Engineering Department, Duke University, Durham, NC 27708 USA (e-mail: vahid.tarokh@duke.edu).

This paper has supplementary downloadable material available at http://ieeexplore.ieee.org., including software implementations (in MATLAB) and real data used in the paper. This material is 20 KB in size.

Digital Object Identifier 10.1109/TSP.2018.2890369

becomes the number of neighbors of the targeted time $t$ for the smoothing step. However, in this work we have removed the smoothing step and apply the multitaper method over the same segment of process, i.e., with the same targeted time $t$ and frequency $w$. With additional smoothness assumptions on the underlying spectra, the authors in [20] have investigated the statistical properties of the spectra estimate.

The paper is organized as follows. In Section II, the evolutionary spectra framework is briefly reviewed. In Section III, the main results are summarized and the key steps of the proofs are presented. In Section IV, the estimate based on the short-time Fourier transform is evaluated through a new time-domain approach, which facilitates the analysis of the estimate in later sections and serves as a much simplified proof compared with the original one. In Section V, the estimate based on the multi-taper method is analyze in the evolutionary spectra framework. In Section VI, a non-parametric stationarity test is proposed and various experimental studies are presented.

### A. Notation

Let $\mathbb{Z}$ and $\mathbb{R}$ denote the set of integers and real numbers, respectively. For integers $a$ and $b$ such that $a < b$, let $[a:b] \triangleq \{a, a+1, \ldots, b\}$. Let $\{X(t)\} \triangleq \{X(t), t \in \mathbb{Z}\}$ denotes a sequence of random variables. $\{X(t)\}$ is called *wide-sense stationary* if its mean is a constant $\mathsf{E}[X(t)] = m_X$ and its auto-covariance depends only on the distance between the time indices $\mathrm{Cov}(X(t), X(s)) = c(t-s)$. Throughout this paper, stationary processes are referred to as wide-sense stationary processes. For two non-negative functions $f(x)$ and $g(x)$, we write $f(x) = O(g(x))$ if there exists some constant $0 < C < \infty$ such that $f(x) \leq Cg(x)$ for sufficiently large $x$. Let $\mathcal{N}(\mu, \sigma^2)$ denote a normal distribution with mean $\mu$ and variance $\sigma^2$. Let $\log(\cdot)$ denote the logarithm function with base 2. For $p \geq 1$, let $\|\cdot\|_p$ denote the $l_p$ norm. We closely follow the notation in [21].

## II. EVOLUTIONARY SPECTRA FRAMEWORK

### A. Brief Review of the framework

In [2], the main focus is the continuous time setting, and the discrete time setting follows immediately. In this work, we will focus on the discrete time setting. In the following, we first briefly review the evolutionary spectra framework. Consider a class of non-stationary processes $\{X(t)\}$, with $\mathsf{E}[X(t)] = 0$ and $\mathsf{E}[X^2(t)] < \infty$ for $t \in \mathbb{Z}$, such that

$$X(t) = \int_{-\pi}^{\pi} \phi_t(w) dZ(w), t \in \mathbb{Z}, \qquad (1)$$

for some family $\mathcal{F}$ of functions $\{\phi_t(w)\}$ (defined on $[-\pi, \pi]$ indexed by $t$) and a measure $\mu(w)$, where $Z(w)$ is an orthogonal increment process with $\mathsf{E}|dZ(w)|^2 = d\mu(w)$. If there exists a family of functions $\{\phi_t(w) = e^{iwt}A_t(w)\}$ such that $\{X(t)\}$ can be represented as in (1) and for any fixed $w$, the Fourier transform of $h_w(t) \triangleq A_t(w)$ (viewed as a function of $t$), denoted by $H_w(v)$, has an absolute maximum at the origin, then $\{X(t)\}$ is called an *oscillatory process* with respect to *oscillatory functions* $\{e^{iwt}A_t(w)\}$, and the evolutionary spectrum at time $t$ with respect to $\mathcal{F}$ is

$$dF_t(w) = |A_t(w)|^2 d\mu(w).$$

*Remark 1:* Note that $h_w(t) \equiv 1$ corresponds to the case when $\{X(t)\}$ is a stationary process, which leads to $H_w(v) \equiv \delta(v)$, where $\delta(\cdot)$ is the Dirac delta function.

Throughout this paper, we assume that $\mu(w)$ is absolutely continuous with respect to Lebesgue measure. Thus the *evolutionary spectral density at time $t$* is

$$f_t(w) = |A_t(w)|^2 \frac{d\mu(w)}{dw}.$$

As mentioned above, for any fixed $w$, $H_w(v)$ is the Fourier transform of $h_w(t)$, i.e.,

$$H_w(v) = \sum_{t=-\infty}^{\infty} h_w(t) e^{-ivt}.$$

Without loss of generality, $A_t(w)$ can be normalized so that for all $w$,

$$A_0(w) = 1, \qquad (2)$$

which implies that $d\mu(w)$ represents the evolutionary spectrum at $t = 0$ and $|A_t(w)|^2$ represents the change relative to $t = 0$. Let

$$B_\mathcal{F}(w) = \int_{-\pi}^{\pi} |v| |H_w(v)| dv,$$

and each family $\mathcal{F}$ of oscillatory functions is called *semi-stationary* if $B_\mathcal{F}(w)$ is bounded for all $w$. Then

$$B_\mathcal{F} = \left( \sup_w B_\mathcal{F}(w) \right)^{-1}$$

is call the *characteristic width of $\mathcal{F}$*. A *semi-stationary process* $\{X(t)\}$ is defined as the one that can be represented as (1) with respect to a semi-stationary family $\mathcal{F}$. Let $\mathcal{C}$ denote the class of semi-stationary families such that $\{X(t)\}$ can be represented as (1). Then

$$B_X = \sup_{\mathcal{F} \in \mathcal{C}} B_\mathcal{F} \qquad (3)$$

is called the *characteristic width of $\{X(t)\}$*. If there exists a family $\mathcal{F}^* \in \mathcal{C}$ with characteristic width equals to $B_X$, $\mathcal{F}^*$ is called the natural representation of $\{X(t)\}$. If there exists no family in $\mathcal{C}$ with characteristic width equals to $B_X$, let $\mathcal{F}^*$ denote any family with characteristic width arbitrarily close to $B_X$. From now on we will only focus on $\mathcal{F}^*$ and the spectral representation with respect to this family. In particular, $A_t(w)$, $d\mu(w)$, and $dF_t(w)$ are all defined with respect to $\mathcal{F}^*$. In this work, we consider that $\{X(t)\}$ admits a natural representation $\mathcal{F}^*$, which implies that $B_{\mathcal{F}^*} = B_X$.

*Remark 2:* It is straightforward to see that semi-stationary processes includes stationary processes as special cases with $\mathcal{F}^* = \{e^{iwt}\}$ and $B_X = \infty$.

The characteristic width of $\{X(t)\}$, $B_X$, can be intuitively viewed as the maximal length of the interval over which the process may be treated as "approximately stationary" [2]. It plays an important role in this framework, however, it is hard to characterize (see Section V-B for a detailed discussion). Priestley [2] proposed a double-window technique to estimate the evolutionary spectral density. The first window is for the short-time Fourier transform and the second window is for smoothing. In this work, however, the second window will be replaced by the multitaper method as smoothing is known to suffer from the bias leakage problem (see Section V for details). The width of the first window $\{g(u), u \in \mathbb{R}\}$ is defined as

$$B_g \triangleq \sum_{u=-\infty}^{\infty} |u||g(u)|.$$

In this work, we focus on *time-limited* windows, i.e., there exists some $L > 0$ such that $g(u) = 0$ for $|u| > L$. Let $G(w)$ denote the Fourier transform of $g(u)$, i.e.,

$$G(w) = \sum_{u=-\infty}^{\infty} g(u) e^{-iuw}. \quad (4)$$

We assume that $g(u)$ is square integrable and without loss of generality it is normalized,

$$2\pi \sum_{u=-\infty}^{\infty} |g(u)|^2 = \int_{-\pi}^{\pi} |G(w)|^2 dw = 1. \quad (5)$$

### B. Uniformly Modulated Processes

It is hard to characterize $B_X$ exactly for semi-stationary processes [22]. However, there is one important class of processes whose characteristic widths can be bounded from below. This class, termed as the *uniformly modulated processes* [2], is of the following form:

$$X(t) = c(t) Y(t), \quad (6)$$

where $Y(t)$ is a stationary process with zero mean and spectral density $f_Y(w)$, and the Fourier transform of $c(t)$ has an absolute maximum at the origin. Thus it follows straightforwardly that

$$X(t) = \int_{-\pi}^{\pi} c(t) e^{iwt} dZ(w),$$

where $\mathsf{E}|dZ(w)|^2 = dF_Y(w)$. The process introduced in (6) is an oscillatory process since $\mathcal{F}_Y = \{c(t) e^{iwt}\}$ is a family of oscillatory functions. The evolutionary spectrum with respect to $\mathcal{F}$ is

$$f_t(w) = c^2(t) f_Y(w).$$

The name, uniformly modulated processes, follows from the fact that for two different frequencies $w_1$ and $w_2$ in $[-\pi, \pi]$, the spectrum is modulated in the same way, i.e.,

$$\frac{f_{t_1}(w_1)}{f_{t_2}(w_1)} = \frac{f_{t_1}(w_2)}{f_{t_2}(w_2)}.$$

From the definition of $B_X$, we have $B_X \geq B_{\mathcal{F}_Y}$.

### III. STATEMENTS OF THE MAIN RESULTS

For semi-stationary processes $\{X(t), 0 \leq t \leq T-1\}$, it is natural to apply the *multitaper method* [18], which identifies $K$ sequences of length $N$ denoted by $\{g_k(u), 1 \leq k \leq K, 1 \leq u \leq N\}$ (assume $N$ to be odd for simplicity of notation). Let $2\pi/N < W < \pi$ denote the frequency resolution of the multitaper method. The details are postponed to Section V. The estimate of the evolutionary spectral density $f_t(w)$ is given as below

$$\hat{f}_t^K(w) = \frac{1}{K} \sum_{k=0}^{K-1} |U_t^{(k)}(w)|^2, \quad (7)$$

where we have

$$U_t^{(k)}(w) = \sum_{u=0}^{T-1} g_k(u-t) X(u) e^{-iwu}$$

with $g_k(u)$ being a set of sequences (shifted so that they are centered around 0) each of length $N < T$ for $-(N-1)/2 \leq u \leq (N-1)/2$ and $1 \leq k \leq K$. Let

$$G_k(\lambda) = \sum_{u=-(N-1)/2}^{(N-1)/2} g_k(u) e^{-i\lambda u}.$$

The expectation of its evolutionary spectral density estimate using the multitaper method is given below.

*Theorem 1:*

$$\mathsf{E}[\hat{f}_t^K(w)]$$
$$= \int_{-\pi}^{\pi} \rho_K(w-\lambda) f_t(\lambda) d\lambda + O\big(B_g^{(K)}/B_X\big)$$
$$= \int_{-\pi}^{\pi} \rho_K(\lambda) f_t(w-\lambda) d\lambda + O\big(B_g^{(K)}/B_X\big),$$

where

$$\rho_K(\lambda) \triangleq \frac{1}{K} \sum_{k=0}^{K-1} |G_k(\lambda)|^2, \quad B_g^{(K)} \triangleq \max_k B_{g_k},$$

and $B_g^{(K)}$ is sufficiently smaller than $B_X$.

Assume that $\|f_t(w)\|_\infty$ is bounded for all $t$, the bias of the estimate can be bounded as follows.

*Theorem 2:* Assume that $\|f_t(w)\|_\infty < \infty$,

$$|\text{Bias}(\hat{f}_t^K(w))| = |\hat{f}_t^K(w) - f_t^K(w)|$$
$$= O\left(\frac{\log N}{K} + W^2 + \frac{B_g^{(K)}}{B_X}\right).$$

When $\{X(t)\}$ is further assumed to be a normal process, the variance of $\hat{f}_t^K(w)$ can be characterized as follows.

*Theorem 3:* Assume that $\|f_t(w)\|_\infty < \infty$ and $\{X(t)\}$ be a normal process,

$$\text{Var}(\hat{f}_t^K(w)) = O\left(\frac{1}{K} + \frac{B_g^{(K)}}{B_X}\right).$$

From Theorem 2 and 3, the mean squared error (MSE) of $\hat{f}_t^K(w)$ is given by the following.

*Corollary 1:* Given the same assumptions as in Theorem 3,

$$\text{MSE}(\hat{f}_t^K(w))$$
$$= O\left(\left(\frac{\log N}{K}\right)^2 + W^4 + \frac{1}{K} + \frac{B_g^{(K)}}{B_X}\right).$$

The proofs of the results are postponed to Section V and appendices and we briefly overview the main ingredients here. Firstly, we analyze $|U(w)|^2$ for general $\{g(u)\}$, which serves as a preliminary estimate (Propostion 1 and Propostion 2) before applying the multitaper method. We take a different approach than Priestley did in [2], in particular, we apply the *pseudo $\delta$-function* argument (see Definition 1) directly in the time domain (Lemma 1) instead of in the frequency domain. This alternative approach allows us to carry out analysis without introducing the *generalized transfer function* (for details see equation (6.6) and Theorem 7.2 in [2]). The benefits of this new approach are twofolds, it makes the variance analysis for the multitaper method straightforward (Theorem 3) and provides a much simplified alternative proof of Propostion 2, which is a slightly different version of Theorem 8.1 in [2], which then leads to Theorem 1. Secondly, by leveraging on a recent approximation result (Theorem 4) on multitaper method by Abreu and Romero [23], we analyze the bias/variance/resolution tradeoffs in the evolutionary spectra framework as in Corollary 1 based on Theorem 2 and Theorem 3.

## IV. APPROXIMATELY UNBIASED ESTIMATE OF THE EVOLUTIONARY SPECTRA

In this section, we start with analyzing a preliminary estimate of the $f_t(w)$. Recall that $g(u)$ is assumed to be a time-limited function. First introduce $J_t(w)$ as follows, for fixed $t \in \mathbb{Z}$ and $w \in [-\pi, \pi]$,

$$J_t(w)$$
$$= \sum_{u=-\infty}^{\infty} g(u-t)X(u)e^{-iwu}$$
$$= \sum_{u=-\infty}^{\infty} g(u-t)\left(\int_{-\pi}^{\pi} A_u(\lambda)e^{i\lambda u}dZ(\lambda)\right)e^{-iwu}$$
$$\stackrel{(a)}{=} \int_{-\pi}^{\pi} \sum_{u=-\infty}^{\infty} g(u-t)A_u(\lambda)e^{-i(w-\lambda)u}dZ(\lambda)$$
$$= \int_{-\pi}^{\pi} e^{-i(w-\lambda)t} \sum_{v=-\infty}^{\infty} g(v)A_{v+t}(\lambda)e^{-i(w-\lambda)v}dZ(\lambda),$$

where the summation and integral can exchange in (a) is because $g(u)$ is a time-limited function.

*Remark 3:* Note that $J_t(w)$ is equivalent to its counterpart $Y_t(w)$ in [2] when $g(u)$ is a symmetric function, i.e., $g(u) = g(-u)$ for all $u$.

In the following, we introduce the pseudo $\delta$-function argument but apply it to the time domain directly. The analysis of the spectra estimate of $f_t(w)$ in [2] depends on an approximation called *pseudo $\delta$-function* and the discrete counterpart can be defined as below. The continuous version can be defined similarly and is used in [2].

*Definition 1:* Consider two functions $a(\cdot): \mathbb{Z} \to \mathbb{R}$ and $b(\cdot): \mathbb{Z} \to \mathbb{R}$. Then $a(u)$ is a pseudo $\delta$-function of order $\epsilon$ with respect to $b(u)$ if, for any $t \in \mathbb{Z}$, there exists $\epsilon$ not depending on $t$ such that

$$\left|\sum_{u=-\infty}^{\infty} a(u)b(u+t) - b(t)\sum_{u=-\infty}^{\infty} a(u)\right| < \epsilon.$$

Now we show the following.

*Lemma 1:* For family $\mathcal{F}^*$, $a(u) \triangleq g(u)e^{-iwu}$ is a pseudo $\delta$-function of $b(u) \triangleq A_u(w)$ with order $O(B_g/B_X)$.

*Proof:* To clarify the role of $u$ as the argument, we write $A_w(u) = A_u(w)$ in this proof. For any $t \in \mathbb{Z}$,

$$\sum_{u=-\infty}^{\infty} g(u)A_w(u+t)e^{-iwu}$$
$$= \sum_{u=-\infty}^{\infty} g(u)A_w(t)e^{-iwu} + R(t)$$

with

$$|R(t)|$$
$$\stackrel{(a)}{=} \frac{1}{2\pi}\left|\sum_{u=-\infty}^{\infty} g(u)e^{-iwu}\int_{-\pi}^{\pi} H_w(v)(e^{iv(t+u)} - e^{ivt})dv\right|$$
$$\leq \frac{1}{2\pi}\sum_{u=-\infty}^{\infty} |g(u)|\int_{-\pi}^{\pi} |H_w(v)||e^{ivu} - 1|dv$$
$$\stackrel{(b)}{\leq} \frac{1}{2\pi}\sum_{u=-\infty}^{\infty} |u||g(u)|\int_{-\pi}^{\pi} |v||H_w(v)|dv$$
$$\stackrel{(c)}{\leq} O(B_g/B_X),$$

where in $(a)$, $A_w(u)$ is substituted by

$$A_w(u) = \frac{1}{2\pi}\int_{-\pi}^{\pi} H_w(v)e^{ivu}dv,$$

$(b)$ follows since $|e^{ix} - 1| \leq |x|$, and $(c)$ follows the definition of $B_g$ and

$$\int_{-\pi}^{\pi} |v||H_w(v)|dv = B_{\mathcal{F}^*}(w) \leq 1/B_X.$$

∎

*Remark 4:* Lemma 1 provides a delta approximation in the time domain directly, unlike the Priestley's frequency domain approach. The new machinery proposed here allows us to carry out the analysis without introducing the *generalized transfer function* [2]. The generalized transfer function also plays a key role in performing the mean and variance analysis (for details of the variance analysis, see Section III in [6]), thus our approach leads to a straightforward analysis of both mean (Proposition 2) and variance (Theorem 3).

From Lemma 1, $J_t(w)$ can be further expressed in the following expression.

*Proposition 1:*

$$J_t(w)$$
$$= \int_{-\pi}^{\pi} e^{-i(w-\lambda)t} A_t(\lambda)\left(\sum_{v=-\infty}^{\infty} g(v)e^{-i(w-\lambda)v}\right)dZ(\lambda)$$
$$+ O(B_g/B_X)\int_{-\pi}^{\pi} e^{-i(w-\lambda)t}dZ(\lambda)$$
$$= \int_{-\pi}^{\pi} A_t(\lambda)G(w-\lambda)e^{-i(w-\lambda)t}dZ(\lambda)$$
$$+ O(B_g/B_X)\int_{-\pi}^{\pi} e^{-i(w-\lambda)t}dZ(\lambda), \qquad (8)$$

where $G(w) = \sum_{v=-\infty}^{+\infty} g(v)e^{-iwv}$.

As one shall see, the relationship between the window choice $g(u)$ and the estimate $|J_t(w)|^2$ of $f_t(w)$ is revealed directly through this time domain approach. This leads to the following proposition.

*Proposition 2:*

$$\mathsf{E}[|J_t(w)|^2]$$
$$= \int_{-\pi}^{\pi} |G(w-\lambda)|^2 f_t(\lambda) d\lambda + O(B_g/B_X) \quad (9)$$
$$= \int_{-\pi}^{\pi} |G(\lambda)|^2 f_t(w-\lambda) d\lambda + O(B_g/B_X). \quad (10)$$

*Proof:* From (8) we have

$$\mathsf{E}[|J_t(w)|^2]$$
$$= \int_{-\pi}^{\pi} |G(w-\lambda)|^2 f_t(\lambda) d\lambda$$
$$+ O(B_g/B_X) \int_{-\pi}^{\pi} A_t(\lambda) G(w-\lambda) d\mu(\lambda) \quad (11)$$
$$+ O(B_g/B_X) \int_{-\pi}^{\pi} \overline{A_t(\lambda) G(w-\lambda)} d\mu(\lambda) \quad (12)$$
$$+ O\left((B_g/B_X)^2\right) \int_{-\pi}^{\pi} d\mu(\lambda). \quad (13)$$

Recall that

$$\int_{-\pi}^{\pi} |G(w)|^2 dw = 1.$$

Now (11) can be bounded as below.

$$\int_{-\pi}^{\pi} A_t(\lambda) G(w-\lambda) d\mu(\lambda)$$
$$\leq \int_{-\pi}^{\pi} |A_t(\lambda)||G(w-\lambda)| d\mu(\lambda)$$
$$\stackrel{(a)}{=} \int_{\Omega} |A_t(\lambda)||G(w-\lambda)| d\mu(\lambda)$$
$$+ \int_{\Omega^c} |A_t(\lambda)||G(w-\lambda)| d\mu(\lambda)$$
$$\leq \int_{-\pi}^{\pi} d\mu(\lambda) + \int_{\Omega^c} |A_t(\lambda)|^2 |G(w-\lambda)|^2 d\mu(\lambda)$$
$$\stackrel{(b)}{<} \infty,$$

where in (a) define $\Omega \triangleq \{w : |A_t(\lambda)||G(w-\lambda)| \leq 1\}$ and $\Omega^c \triangleq [-\pi, \pi]/\Omega$, (b) follows because of (5) and

$$\int_{\Omega^c} |A_t(\lambda)|^2 |G(w-\lambda)|^2 d\mu(\lambda)$$
$$\leq \int_{-\pi}^{\pi} |G(w-\lambda)|^2 f_t(\lambda) d\lambda \leq \int_{-\pi}^{\pi} f_t(\lambda) d\lambda.$$

Since (12) and (13) can be bounded similarly, this ends the proof for (9). Observe that both $|G(w)|^2$ and $f_t(w)$ for fixed $t$ are periodic functions with period $2\pi$, thus (10) follows from (9). ∎

Given a sample record $\{X(0), X(1), ..., X(T-1)\}$ of length $T$, for $0 \leq t \leq T-1$, let

$$U_t(w) = \sum_{u=0}^{T-1} g(u-t) X(u) e^{-iwu}. \quad (14)$$

If we have that $B_g$ is sufficiently smaller than $B_X$ and $B_X$ sufficiently smaller than $T$, then for $t$ large enough, $U_t(w)$ becomes almost identical to $J_t(w)$ and the end effects are negligible. This holds since we are dealing with $g(u)$ that is time-limited, i.e., $g(u) = 0$ for $|u| > N$ for some $N$. Thus we have for $N/2 < t < T - N/2 - 1$,

$$\mathsf{E}[|U_t(w)|^2] = \int_{-\pi}^{\pi} |G(w-\lambda)|^2 f_t(\lambda) d\lambda + O(B_g/B_X).$$

To understand the impact of $|G(w)|^2$ on the bias of the estimate $|U_t(w)|^2$, consider the ideal case where $|G(w)|^2 = \delta(w)$. Then since $|G(w)|^2$ is normalized, we have

$$\mathsf{E}[|U_t(w)|^2] = f_t(w) + O(B_g/B_X).$$

Thus in this case $|U_t(w)|^2$ becomes an unbiased estimate of $f_t(w)$ up to $O(B_g/B_X)$. Intuitively speaking, the bias is controlled by the sidelobe of $|G(w)|^2$: less sidelobe would lead to a less biased estimate. To quantify the relationship, definitions similar to $B_g$ and $B_X$ in the frequency domain are needed. Let

$$\tilde{B}_g \triangleq \int_{-\pi}^{\pi} w|G(w)|^2 dw, \quad \tilde{B}_\mathcal{F}(t) \triangleq \int_{-\pi}^{\pi} |\lambda||L_t(\lambda)|d\lambda,$$

where $L_t(\lambda) \triangleq \int_{-\pi}^{\pi} f_t(v) e^{-i\lambda v} dv$. The minimum "width" of $f_t(w)$ for any $t$ is thus $\tilde{B}_X \triangleq \sup_{\mathcal{F} \in \mathcal{C}} \tilde{B}_\mathcal{F}$ with $\tilde{B}_\mathcal{F} \triangleq \left(\sup_t \tilde{B}_\mathcal{F}(t)\right)^{-1}$.

Roughly speaking, $\tilde{B}_g$ and $\tilde{B}_X$ characterize the bandwidth of $g(u)$ and $f_t(w)$ in the frequency domain, respectively. Then in order to estimate $f_t(w)$, it has to be changing more slowly than $|G(w)|^2$ for each $t$. The following lemma is immediate and the proof can be found in Appendix A.

*Lemma 2:* $|G(w)|^2$ is a pseudo $\delta$-function of $f_t(w)$ for each $t$ with order $O(\tilde{B}_g/\tilde{B}_X)$.

Together with (5), Lemma 2 leads to

$$\mathsf{E}[|U_t(w)|^2] = f_t(w) + O(\tilde{B}_g/\tilde{B}_X) + O(B_g/B_X).$$

Therefore $|U_t(w)|^2$ is an unbiased estimate up to approximations in both time and frequency domain.

For stationary processes, it is well-known that simple periodogram type of estimate (as in (14) but without $\{g(u)\}$) has a mean involving Fejér's kernel and is not a satisfactory estimate [21], [17]. An excellent numerical example for an $AR(4)$ model is provided in [17, Chapter 7.1]. In the asymptotic regime, either smoothing or tapering is needed to obtain a consistent estimate; while in the non-asymptotic regime, there is significant bias leakage because of the sidelobes of the Fejér kernel [17]. Different tapering techniques have been developed over the years and the *multitaper method* by Thomson [18] is the most widely used technique to reduce both the bias leakage and variance of the estimate. In the evolutionary spectra framework, however, additional constraint $O(B_g/B_X)$ is crucial to the performance of the estimate. Therefore the non-stationarity plays an important role in the bias/variance/resolution tradeoff as we shall see in the next section.

## V. Estimate Based on the Multitaper Method

Thomson's multitaper method [18] has been widely applied to various fields including wireless communnincation [24], neuroscience [25], climate science [26]. In a recent paper by Abreu and Romero [23], the authors provide a rigorous proof of an important heuristic discovered by Thomson. They show that the averaged taper is close to an ideal band-pass filter over $[-W, W]$, i.e., $(1/2W)\mathbf{1}_{[-W,W]}$ in $L_1$ distance. First we briefly review Thomson's multitaper method [18] and the discrete prolate spheroidal sequences (DPSS) or Slepian sequences [17], [27]–[29].

### A. Thomson's Multitaper Method

Consider $N$ sample records $\{X(0), \ldots, X(N-1)\}$.[1] Assume that the sampling frequency is 1, then for a sequence of length $N$, the fundamental frequency is $2\pi/N$ and the Nyquist frequency is $\pi$. For $2\pi/N < W < \pi$, one wishes to find sequences with spectral densities concentrated over $[-W, W]$. We will refer to $W$ as the resolution of the estimate. This problem was first investigated in a series of papers by Slepian, Laudau, Pollak [27]–[29]. The solution turns out to be a set of sequences $v_k(N, W; u)$, $0 \leq u \leq N-1$, $0 \leq k \leq N-1$, which satisfy the following eigenvalue equation

$$\sum_{u'=0}^{N-1} \frac{\sin W(u-u')}{\sin \pi(u-u')} v_k(N, W; u')$$
$$= \lambda_k(N, W) v_k(N, W; u).$$

These $N$ eigenvectors $v_k(N, W; \cdot)$ are called the discrete prolate spheroidal and they are ordered by their eigenvalues $1 > \lambda_0(N, W) > \lambda_1(N, W) > \cdots > \lambda_{N-1}(N, W) > 0$. It is well-known that the first $K = \lfloor 2NW/2\pi \rfloor = \lfloor NW/\pi \rfloor$ eigenvalues are close to 1.

*Remark 5:* The choice of using $N$ as the length of the sample records instead of $T$ is on purpose. $T$ is the length of the whole sample records, while $N$ will be used as the length of a time-limited function $g(u)$ discussed later in this section. If the process is indeed stationary, one would choose $T = N$.

The discrete prolate spheroidal wave functions are denoted by $V_k(N, W; \lambda)$ for $1 \leq k \leq K$, where

$$V_k(N, W; \lambda) = (-1)^k \epsilon_k \sum_{u=0}^{N-1} v_k(N, W; u) e^{-i\lambda(u-(N-1)/2)},$$

where $\epsilon_k = 1$ when $k$ is even and $\epsilon_k = \sqrt{-1}$ when $k$ is odd. For simplicity of notation, we suppress $N$ and $W$ and write $v_k(u) = v_k(N, W; u)$, $V_k(\lambda) = V_k(N, W; \lambda)$, and $\lambda_k = \lambda_k(N, K)$. These $K$ functions satisfy two types of orthogonality over $[-W, W]$ and $[-\pi, \pi]$, respectively

$$\int_{-W}^{W} V_k(\lambda) \overline{V_l(\lambda)} d\lambda = \begin{cases} \lambda_k, & \text{for } k = l; \\ 0, & \text{otherwise}. \end{cases}$$

$$\int_{-\pi}^{\pi} V_k(\lambda) \overline{V_l(\lambda)} d\lambda = \begin{cases} 1, & \text{for } k = l; \\ 0, & \text{otherwise}. \end{cases} \quad (15)$$

Observe that $|V_k(\lambda)|^2$ can be rewritten as follows,

$$|V_k(\lambda)|^2 = \left| \sum_{u=0}^{N-1} v_k(u) e^{-i\lambda u} \right|^2. \quad (16)$$

Consider the average of the $K$ tapered estimates,

$$\rho_K(\lambda) \triangleq \frac{1}{K} \sum_{k=0}^{K-1} |V_k(\lambda)|^2. \quad (17)$$

[1] These $N$ sample records are a consecutive subsequence from the $\{X(0), \ldots, X(T-1)\}$.

It has been observed numerically that $\rho_K(\lambda)$ is close to $(1/2W)\mathbf{1}_{[-W,W]}(\cdot)$ by Thomson [18], which is justified recently by Abreu and Romero [23] as given below.

*Theorem 4 ([23]).* Let $N \geq 2$ denote the length of the sequence, $2\pi/N < W < \pi$ and set $K = \lfloor NW/\pi \rfloor$. Then

$$\left\| \rho_K(\cdot) - \frac{1}{2W} \mathbf{1}_{[-W,W]}(\cdot) \right\|_1 = O\left( \frac{\log N}{K} \right).$$

In the following section, we apply this result to analyze the performance of the multitaper method for semi-stationary processes.

### B. Estimate of the Evolutionary Spectra Based on the Multitaper Method

For stationary processes, the bias and variance of the multitaper spectral estimate (17) has been investigated [30], [31], [32]. In this section, we investigate its performance for semi-stationary processes. Let $g(u)$ be a time-limited function, i.e.,

$$|g(u)| = 0, \text{ for } |u| > (N-1)/2,$$

where $N$ is assumed to be odd. Apply the multitaper method on $\{X(t), 0 \leq t \leq T-1\}$ with

$$g_k(u) \triangleq v_k(u + (N-1)/2) \text{ for } 0 \leq k \leq K-1,$$

then for $t > (N-1)/2$ we have

$$U_t^{(k)}(w) = \sum_{u=0}^{T-1} g_k(u-t) X(u) e^{-iwu}$$
$$= \sum_{u=t-(N-1)/2}^{t+(N-1)/2} g_k(u-t) X(u) e^{-iwu}.$$

From Proposition 2 and (16),

$$\mathsf{E}[|U_t^{(k)}(w)|^2]$$
$$= \int_{-\pi}^{\pi} |G_k(w-\lambda)|^2 f_t(\lambda) d\lambda + O(B_{g_k}/B_X),$$

where

$$G_k(\lambda) = \sum_{u=-\infty}^{\infty} g_k(u) e^{-i\lambda u} = \sum_{u=-(N-1)/2}^{(N-1)/2} g_k(u) e^{-i\lambda u}.$$

The estimate of $f_t(w)$ is the average of $|U_t^{(k)}(w)|^2$,

$$\hat{f}_t^K(w) = \frac{1}{K} \sum_{k=0}^{K-1} |U^{(k)}(w)|^2$$

and the mean of the estimate given in Theorem 1.

There is a bias/variance/resolution tradeoff for the estimate $\hat{f}_t^K(w)$. Assuming that $\|f_t(w)\|_\infty$ is bounded for all $t$, Theorem 2 can be proved by invoking Theorem 4 as given below.

*Proof:* First, the bias can be bounded,

$$|\text{Bias}(\hat{f}_t^K(w))|$$
$$= |\mathsf{E}[\hat{f}_t^K(w)] - f_t(w)|$$
$$\leq \left| \int_{-\pi}^{\pi} \rho_K(\lambda) f_t(w-\lambda) d\lambda - f_t(w) \right|$$
$$+ O(B_g^{(K)}/B_X).$$

From Theorem 4,

$$\left|\int_{-\pi}^{\pi} \rho_K(\lambda) f_t(w-\lambda) d\lambda - f_t(w)\right|$$

$$\leq \left|\int_{-\pi}^{\pi} \left(\rho_K(\lambda) - \frac{1}{2W}\mathbf{1}_{[-W,W]}(\lambda)\right) f_t(w-\lambda) d\lambda\right|$$

$$+ \left|\int_{-\pi}^{\pi} \frac{1}{2W}\mathbf{1}_{[-W,W]}(\lambda) f_t(w-\lambda) d\lambda - f_t(w)\right|$$

$$\overset{(a)}{\leq} C(\log(N)/K + W^2),$$

where $(a)$ follows from Theorem 4 and the assumption that $\|f_t(w)\|_\infty < \infty$ and $C$ is some positive constant. ∎

When $X(t)$ is a normal process, the variance of $\hat{f}_t^K(w)$ can be characterized as in Theorem 3 and the proof can be found in Appendix B. Now the MSE of $\hat{f}_t^K(w)$ can be bounded as shown in Corollary 1,

$$O\left(\left(\frac{\log N}{K}\right)^2 + W^4 + \frac{1}{K} + \frac{B_g^{(K)}}{B_X}\right),$$

where we have used the fact that the cross term $W^2(\log N/K)$ is dominated by either $W^2$ or $(\log N/K)$ depending on whether $W^2 \geq (\log N/K)$ or $W^2 \leq (\log N/K)$, and this argument applies to all the other cross terms as well.

For the uniformly modulated processes $X(t) = c(t)Y(t)$, where $Y(t)$ is a stationary process as defined in Section II-B, $B_X$ can be lower bounded by $B_{\mathcal{F}_Y}$. Thus, based on Corollary 1, the MSE of the estimate in the case can be further bounded as

$$\left(\frac{\log N}{K}\right)^2 + W^4 + \frac{1}{K} + \frac{B_g^{(K)}}{B_{\mathcal{F}_Y}}. \tag{18}$$

Recall that $K = \lfloor NW/\pi \rfloor$ and $W$ can be as small as $2\pi/N$. Thus for stationary processes, the MSE of the spectral density estimate decreases as $N$ grows. However, this is no longer the case for the semi-stationary processes as $B_g^{(K)}/B_X$ may become the dominant term for $N$ large enough. We demonstrate this point through the following example appeared in [2], [6], [7].

**Example.** Consider the following semi-stationary process

$$X(t) = c(t)Y(t), \text{ for } 1 \leq t \leq T$$

where $a = 200$, $c(t) = e^{(t-T/2)^2/2a^2}$, and

$$Y_t = 0.8Y_{t-1} - 0.4Y_{t-2} + Z_t,$$

with $Z_t \sim \mathcal{N}(0, 100^2)$. It is shown in [6] that $B_{\mathcal{F}_Y} = a\sqrt{\pi/2}$. We now use this fact to evaluate (18) and compare it with the MSE without considering non-stationarity, i.e.,

$$\left(\frac{\log N}{K}\right)^2 + W^4 + \frac{1}{K}. \tag{19}$$

In Fig. 1, we compare the MSE in these two cases and the parameter $K$ is optimized over $\{2, 3, \ldots, N-1\}$ since $2\pi/N < W < \pi$. In Fig. 1, for each $N$, the MSE are computed with respect to the optimal $K$. We can see that the MSE in (18) starts to increase for larger $N$, i.e., larger $N$ will no longer be beneficial for estimating the evolutionary spectra of semi-stationary processes. This is because larger $N$ does not provide more information of the spectra since it is changing over time. This is in contrast with the stationary case, where larger $N$ would improve the performance of the estimate. The relationship between the $N$ and the corresponding optimal $K$ is shown in Fig. 2. Note that (19) is derived in [23] and the optimal $K$ scales roughly as $N^{-4/5}$. For each $N$, the $K$ that minimize the MSE do not vary much in both cases.

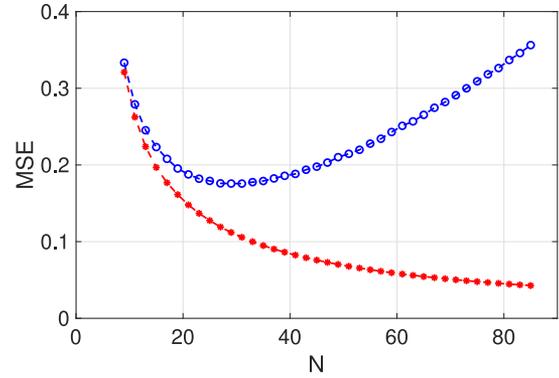

Fig. 1. The blue circle line corresponds to the MSE in (18) with respect to the optimal $K$ for each $N$ and the red dot line corresponds to the MSE in (19) with respect to the optimal $K$ for each $N$.

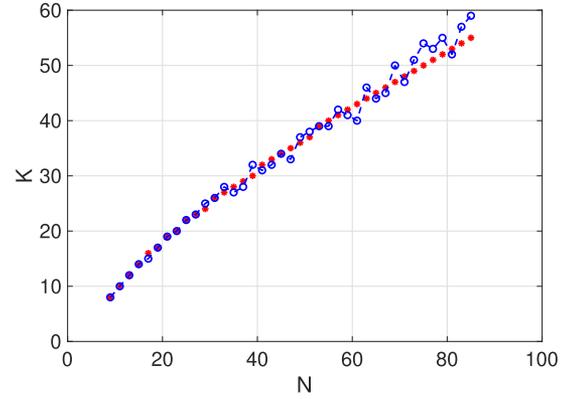

Fig. 2. The blue circle line is the relationship between $N$ and the corresponding optimal $K$ evaluated according to (18) and the red dot line is the relationship between $N$ and the corresponding optimal $K$ evaluated according to (19).

For non-stationary processes, it is a natural idea to approximate it by stationary processes locally. Heuristic methods, such as segmentation, have been developed [33] to deal with non-stationarity. In the evolutionary spectra framework, $B_X$ defined in (3) can be roughly interpreted the longest "approximately stationary" segment [2] of the semi-stationary process. It is thus tempting to estimate $B_X$. However, it seems that $B_X$ is more of a theoretical technique rather than providing fundamental meanings. Its definition is tailored to get the first order approximation of the estimate, which can be partially seen from Section IV. Furthermore, characterizing $B_X$ is highly non-trivial as shown by Mélard in [22]. As a comparison, in the locally stationary processes framework [4], the authors characterized the optimal choice of $N$ as $N_{\text{opt}}$, the length of the stationary segment [34] through minimizing the MSE of a local covariance estimate. While the characterization is interesting from the theoretical point of view, its application is limited due to its dependence on the true unknown parameters.

As a natural application of the evolutionary spectral density estimate, we propose a non-parametric stationarity test in the next Section.

## VI. STATIONARITY TEST

The evolutionary spectral density estimate suggests a statistical test for the stationarity of a process, as first discussed in Priestley's paper [2] and later investigated by Priestley and Subba Rao (PSR test) in [7]. The original version of the PSR test uses the smoothing technique by introducing a second window, which suffers from bias leakage problems as discussed in Section IV. In a recent package developed by Constantine and Percival [35], smoothing is replaced by the multitaper method. This modified PSR test has been served as a baseline to when compared with other stationarity tests, e.g., in [36]. Based on the results from Section V-B, we attempt to provide some insights into the choice of the parameters in the test. Furthermore, a non-parametric version of the stationarity test is proposed, which is based on the Friedman test [37], [38] and is robust to the underlying distribution. It serves as a complementary test to the existing stationarity tests, in the sense that it is more conservative than PSR, see Section VI-C for details.

### A. PSR Stationarity Test With the Multitaper Method

Let $f_t(w)$ denote the evolutionary spectral density of a semistationary process $\{X(t) : 0 \leq t \leq T-1\}$. Consider the estimate $\hat{f}_t^K(w)$ based on the multitaper method as in Section V and recall that

$$\hat{f}_t^K(w) = \frac{1}{K} \sum_{k=0}^{K-1} |U_t^{(k)}(w)|^2,$$

where

$$U_t^{(k)}(w) = \sum_{u=t-(N-1)/2}^{t+(N-1)/2} g_k(u-t)X(u)e^{-iwu}.$$

It is a common practice to take the logarithm of the estimate, which stabilizes its variance [39]. Let

$$Y_{ij} = \log \hat{f}_{t_i}^K(w_j),$$

Moreover, to apply the two-way analysis of variance (ANOVA) test [40], it has to be assumed that the distribution of $\log \hat{f}_t^K(w)$ is approximately normal [7]. More specifically, it can be shown that $W_{ij} = Y_{ij} - \psi(K) + \log(K)$ is approximately distributed according to the normal distribution with mean 0 and variance $\sigma^2 = \psi'(K)$ for $K \geq 5$, where $\psi(\cdot)$ and $\psi'(\cdot)$ denote the digamma function and the trigamma function, respectively (see [41, Section II] and [42] for details).

The approximate independence in time is obtained by choosing non-overlapping short windows of length $N$ and the approximate independence in frequency is by choosing frequencies that are $2\pi(K+1)/(N+1)$ apart.[2] Now the problem reduces to a two-way ANOVA test for $W_{ij}$ for $i \in [1:I]$ and $j \in [1:J]$,

where $I = \lfloor T/N \rfloor$ and $J$ is the number of frequencies chosen $2\pi(K+1)/(N+1)$ apart. Let

$$W_{..} = (1/IJ) \sum_{i=1}^{I} \sum_{j=1}^{J} W_{ij},$$

$$W_{i\cdot} = (1/J) \sum_{j=1}^{J} W_{ij},$$

$$W_{\cdot j} = (1/I) \sum_{i=1}^{I} W_{ij}.$$

Between times variance with degrees of freedom $I-1$ concerns how uniform are $\{W_{ij}\}$ over the time indices $1 \leq i \leq I$,

$$S_T = J \sum_{i=1}^{I} (W_{i\cdot} - W_{..})^2.$$

Similarly, between frequencies variance with degrees of freedom $J-1$ is

$$S_F = I \sum_{j=1}^{J} (W_{\cdot j} - W_{..})^2.$$

Interaction and residual variance with degrees of freedom $(I-1)(J-1)$ is

$$S_{I+R} = \sum_{i=1}^{I} \sum_{j=1}^{J} (W_{ij} - W_{i\cdot} - W_{\cdot j} + W_{..})^2.$$

The null hypothesis is that the process is stationary and the alternative hypothesis is that the process is non-stationary. The test steps are described in the following.

1) First, test the interaction and residual sum of squares using $S_{I+R}/\sigma^2 = \chi^2_{(I-1)(J-1)}$.
2) If the interaction and residual is not significant, one conclude that the process is a uniformly modulated process.[3] Then proceed to test $S_T/\sigma^2 = \chi^2_{(I-1)}$. If the between-times is not significant, conclude that the process is stationary. Otherwise, conclude that the process is non-stationary.
3) If the interaction and residual is significant, conclude that the process is non-stationary.

### B. A Non-Parametric Stationarity Test

There are two main assumptions of the two-way ANOVA test: (1) the samples are uncorrelated and (2) the residuals are normally distributed. There has been extensive research on the robustness of the assumptions for ANOVA test. It is known that the test statistics depend heavily on the first assumption and is less sensitive to the second assumption. The latter is shown empirically first in [43] and later in [44].

More specifically, the degree of violation of the normal distribution is usually characterized by the skewness $\beta_1$ and flatness $\beta_2$ of the distribution, where $\beta_1 = \mathsf{E}[(X-\mu)^3]/\sigma^3$ and $\beta_2 = \mathsf{E}[(X-\mu)^4]/\delta^4$, where $\mu$ and $\delta^2$ denote the mean and variance of $X$, respectively. The test statistics are less sensitive to the skewness and flatness of $X$, essentially due to the central limit theorem as the the test statistics are based on summation of many terms. In the PSR test, the test results are more reliable when the degrees of freedom of time and frequency are large. On the other hand, nonparametric test, e.g., the rank-based Friedman test [37], [38], has an edge when the number of test samples is relatively small.

---

[2]Buffers are needed at the beginning and at the end when sample in frequency to overcome the edge effect. The size of the buffer could be chosen from $[B/2, B]$, where $B \triangleq 2\pi(K+1)/(N+1)$.

[3]The test of the sum of interaction and residual being not significant implies that the time and frequency can be decoupled and thus the process can be expressed in the form of uniformly modulated process. See [7] for details.

We now propose the non-parametric test, which will be referred to as *rank-based stationarity test* or *RS test* in short. Take $\{W_{ij}\}$ introduced in the previous section. In the time-frequency table filled by $\{W_{ij}\}$, rank the elements in each column in an increasing order (i.e., 1 corresponds to the smallest element) to form a table of ranks: $\{R_{ij}\}$. Whenever there is a tie among $k$ elements in the same column, assign the mean rank of the $k$ elements. Similar to the two-way ANOVA test, let $R_{..}$ denote the mean rank of all ranks, denote $R_{i.}$ the mean rank of row $i$. The sum of square of ranks $SS_R$ is $SS_R = J \sum_i (R_{i.} - R_{..})^2$. The test statistics $t_R = SS_R/\text{const}$, where $\text{const} = I(I+1)/12$. It is known that $t_R$ is (approximately) distributed according to $\chi^2_{I-1}$ [37], [38].

*Remark 6:* Conventionally, the rows are ranked and then the ranks in each column are summed up to form the test statistic. To be consistent with the two-way ANOVA test, the role of row and column are switched in this work.

### C. Simulations

In this section, the performance of the proposed non-parametric stationarity test is evaluated and compared with the PSR test for a variety of synthetic data and real data. In our simulation, we use the multitaper function *pmtm* in MATLAB (R2016a) with default values as in [35]: number of tapers is 5, number of non-overlapping blocks is $\max\{2, \log(T)\}$, and buffer size $0.7B$ where $B = 2\pi(K+1)/(N+1)$. The number of tapers is much smaller than the one we used for estimating the evolutionary spectra in Section V-B.

Before the simulations, we first try to justify the choice of number of tapers. One key factor is that testing stationarity is a task of different nature compared with estimating the underlying evolutionary spectra. In particular, testing stationarity requires a good amount of "independent" samples in both time and frequency domain. Recall that this rough independence between the samples is achieved by sampling in non-overlapping blocks as well as sampling frequencies that are $2\pi(K+1)/(N+1)$ apart. Thus there is a hidden penalty for choosing large $K$ (but less than $2NW$), because large $K$ will reduce the number of samples for performing the hypothesis test. Therefore, instead of solely focusing on MSE, it is reasonable to add to it a penalty $c(K)$ that is an increasing function of $K$. For simplicity of demonstration, we choose $c(K) = K$ and evaluate the example in Section V based on the summation of MSE in (18) and $K$ in Fig. 3. Here $N$ is chosen to be $T/\max\{2, \log(T)\}$ with $T = 512$. Thus number of chosen tapers should be 4 or 5. A thorough study of this penalized approach is out of the scope of this paper and will be pursued in our future works. Note that window design (choosing weights and length of the window) is hard even for stationary processes and "in practice it is advisable to experiment with a range of windows" (see Section 10.4 in [21] for details).

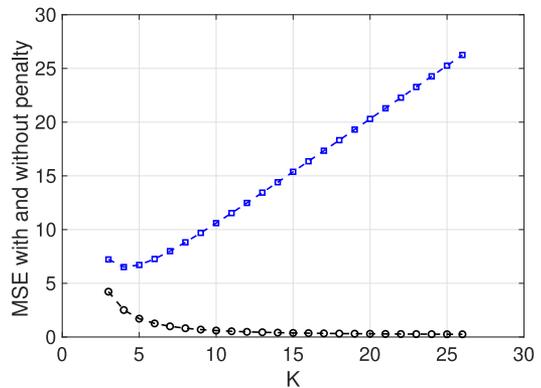

Fig. 3. The blue square line corresponds to the MSE in (18) plus a penalty term $K$ and the black circle line corresponds to the MSE in (18).

TABLE I
EMPIRICAL SIZE COMPARISON (%)

| models | PSR  | RS  |
|--------|------|-----|
| (a)    | 11.1 | 1.9 |
| (b)    | 17.7 | 2.9 |
| (c)    | 11.2 | 2.7 |
| (d)    | 12.7 | 2.5 |
| (e)    | 14.8 | 2.9 |
| (f)    | 15.3 | 2.6 |
| (g)    | 78.7 | 6.1 |

*1) Synthetic Data:* The performance of a test is evaluated based on its empirical *size* and *power* values. Generate $M = 1000$ sample paths/realizations each with length $T = 512$ and let the nominal size of the test be 0.05. The null hypothesis $H_0$ is that the process is stationary and the alternative hypothesis $H_1$ is that the process is not stationarity.

For the size comparison, we generate sample paths from various stationary processes and count the number of rejections of the null hypothesis. Consider the following set of stationary autoregressive and moving-average (ARMA) models used in [45] The noise term $Z(t)$ is distributed according to $\mathcal{N}(0,1)$.
  a) i.i.d. standard normal
  b) AR(1): $X(t) = 0.9X(t-1) + Z(t)$.
  c) AR(1): $X(t) = -0.9X(t-1) + Z(t)$.
  d) MA(1): $X(t) = Z(t) + 0.8Z(t-1)$.
  e) MA(1): $X(t) = Z(t) - 0.8Z(t-1)$.
  f) ARMA(1, 2): $X(t) = -0.4X(t) + Z(t) - 0.8Z(t-1)$.
  g) AR(2): $X(t) = \alpha_1 X(t-1) + \alpha_2 X(t-2) + Z(t)$ with $\alpha_1 = 1.385929$ and $\alpha_2 = -0.9604$ (from [46]).

The empirical sizes for PSR is smaller in [45] than that in Table I, but still at least twice as large as the empirical sizes of RS for all the models in Table I. The differences in PSR from [45] could be due to differences in parameters such as number of tapers, number of non-overlapping blocks, buffer size, etc.

For the power comparison, we generate sample paths from semi-stationary processes and count the number of acceptances of the null hypothesis. We focus on the uniformly modulated processes as in [2], [7]. As in Section V-B, we focus on the following model,

$$X(t) = e^{(t-T/2)^2/2a^2} Y(t), \qquad (20)$$

where $a = 200$ and $Y_t = 0.8Y_{t-1} - 0.4Y_{t-2} + Z_t$ with $Z_t \sim \mathcal{N}(0, 100^2)$. For all the models from Table I, generate uniformly modulated processes by multiplying each of them with $e^{(t-T/2)^2/2a^2}$. To make the numbering consistent with Table I, these models are also numbered from $(a)$ to $(g)$ and model (20) will be numbered as $(h)$ in the table below.

TABLE II
EMPIRICAL POWER COMPARISON (%)

| models | PSR | RS |
|---|---|---|
| (a) | 96.7 | 88.4 |
| (b) | 96.8 | 82.5 |
| (c) | 97.3 | 88.9 |
| (d) | 96.4 | 88.1 |
| (e) | 97.3 | 87.1 |
| (f) | 97.1 | 86.2 |
| (g) | 98.3 | 76.7 |
| (h) | 96.4 | 88.4 |

Since the empirical size of RS is smaller than that of PSR but the empirical power is also smaller, RS is a more conservative test compared with PSR.

*2) Real Data:* We consider a real data example called *ecgrr* used in [35] and comes from 'RR interval time series modeling: A challenge from PhysioNet and Computers in Cardiology 2002' site of PhysioNet. The data are the RR intervals (beat-to-beat intervals measured between successive peaks of the QRS complex) for patients in normal sinus rhythm (record 16265 of the MIT-BIH database). The length of the sample $T = 512$ and the nominal size of the test be 0.05 as in the previous section. Recall that the test statistics for PSR are $S_T/\sigma^2$ and $S_{I+R}/\sigma^2$ and that for RS is $t_R$. Both PSR and RS suggest that the the process is non-stationary and their test statistics are summarized in the following tables.

TABLE III
TEST RESULT OF PSR

| | $S_T/\sigma^2$ | $S_{I+R}/\sigma^2$ |
|---|---|---|
| Chi-square quantile | 16.919 | 72.1532 |
| PSR | 40.4927 | 67.7689 |

TABLE IV
TEST RESULT OF RS

| | $t_R$ |
|---|---|
| Chi-square quantile | 16.919 |
| RS | 22.8 |

*3) Real Data: Purchasing Power Parity:* In economics, a common practice to test the purchasing power parity (PPP) hypothesis via testing the stationarity of real exchange rates (RER). Some earlier studies using unit root tests yielded results that were not favorable to PPP (see for example [47]–[49]). In this real data study, we test the stationarity of RER of four countries (Canada, China, Japan, UK) with respect to US over the period of January 1970 to December 2017. The monthly data of RER were calculated by $E \cdot P^*/P$, where $E$, $P^*$ and $P$ respectively denote the nominal exchange rates, the foreign price level (evaluated using consumer price index) and the domestic price level, using data sources from International Financial Statistics of the International Monetary Fund, Financial Statistics of the Federal Reserve Board, Haver Analytics, and the Pacific Exchange Rate Service. In the experiments, we take the widely used transform: the log first-order difference, i.e., $\log(X_t) - \log(X_{t-1})$ for $2 \leq t \leq N$ with $\{X_t : 1 \leq N\}$ denote the RER.

TABLE V
TEST RESULTS OF PSR

| | $S_T/\sigma^2$ | $S_{I+R}/\sigma^2$ |
|---|---|---|
| Chi-square quantile (5%) | 15.5073 | 36.415 |
| Chi-square quantile (1%) | 20.0902 | 42.9798 |
| PSR (Canada) | 69.4236 | 42.6526 |
| PSR (China) | 637.9477 | 29.5066 |
| PSR (Japan) | 18.4479 | 13.6801 |
| PSR (UK) | 24.1467 | 32.9519 |

Based on RS test and $S_T/\sigma^2$ of the PSR test, one could order the stationarity of the four countries (in terms of RER) from the most stationary to least stationary as: Japan, UK, Canada, China. When choosing 1% as the nominal size of the test, RS test suggests that RER of both Japan and UK are stationary with Canada close to stationary; while PSR test suggests that RER of Japan is stationary and RER of UK is close to stationary.

TABLE VI
TESTS RESULT OF RS

| | $t_R$ |
|---|---|
| Chi-square quantile (5%) | 15.5073 |
| Chi-square quantile (1%) | 20.0902 |
| RS (Canada) | 22.0 |
| RS (China) | 30.8 |
| RS (Japan) | 16.3 |
| RS (UK) | 19.7 |

Both tests (under both 5% and 1%) suggest that RER of China is non-stationary.

## VII. CONCLUDING REMARKS

In this work, we investigate the spectrum estimation for an important class of non-stationary processes developed by Priestley. We propose and analyze an improved estimate within the evolutionary spectra framework. The analysis is based on a novel alternative delta approximation in the time domain, as well as leveraging on a recent concentration result on the multitaper method. The estimate is then applied to develop a non-parametric stationarity test, which is complementary to the existing stationarity tests. There are several interesting future directions. For parameter design, the penalized approach to understand the parameter design for stationarity tests will be pursued. Regarding the computational constraints for spectra estimation, which is crucial for very long time series, it will be interesting to incorporate the SLEX based methods into the framework to handle the computational issues.

## ACKNOWLEDGMENT

The authors would like to thank the anonymous reviewers for their valuable comments and suggestions to improve the quality of the paper. The first author would like to thank Luís Daniel Abreu and José Luis Romero for clarifying proofs in [23].

## APPENDIX A
## PROOF OF LEMMA 1

*Proof:* The statement can be proved by observing

$$\int_{-\pi}^{\pi} |G(w)|^2 f_t(w+v) dw = f_t(v) \int_{-\pi}^{\pi} |G(w)|^2 dw + R(v),$$

where

$$|R(v)| = \left| \int_{-\pi}^{\pi} w |G(w)|^2 f_t'(v + \eta(w)w) dw \right|$$

$$\leq \sup_{-\pi \leq w \leq \pi} |f_t'(w)| \int_{-\pi}^{\pi} |w| |G(w)|^2 dw,$$

where $0 \leq \eta(w) \leq 1$ for all $w$. ∎

## APPENDIX B
## PROOF OF THEOREM 3

*Proof:* The variance of $\hat{f}_t^K(w)$ can be expressed as,

$$\text{Var}(\hat{f}_t^K(w))$$

$$= \text{Var}\left( \frac{1}{K} \sum_{k=0}^{K-1} |U_t^{(k)}(w)|^2 \right)$$

$$= \frac{1}{K^2} \sum_{k,l=0}^{K-1} \text{Cov}\left( |U_t^{(k)}(w)|^2, |U_t^{(l)}(w)|^2 \right).$$

First, $\text{Cov}\left( |U_t^{(k)}(w)|^2, |U_t^{(l)}(w)|^2 \right)$ can be rewritten as (21), which is shown at the bottom of this page. Similar to Proposition 2, we can further express $S_1$ and $S_2$ as (22) and (23), respectively. Now, since $\|f_t(w)\|_\infty < \infty$,

$$\left| \sum_{k,l=0}^{K-1} \text{Cov}\left( |U_t^{(k)}(w)|^2, |U_t^{(l)}(w)|^2 \right) \right|$$

$$\leq \left| \sum_{k,l=0}^{K-1} S_1 \right| + \left| \sum_{k,l=0}^{K-1} S_2 \right|$$

$$\stackrel{(a)}{\leq} 2K + \sum_{k=0}^{K-1} O(B_{g_k}/B_X),$$

---

$$\text{Cov}\left( |U_t^{(k)}(w)|^2, |U_t^{(l)}(w)|^2 \right)$$

$$= \text{Cov}\left( \sum_{u_1,u_2} g_k(u_1-t)g_k(u_2-t)X(u_1)X(u_2)e^{-iwu_1}e^{iwu_2}, \sum_{u_3,u_4} g_l(u_3-t)g_l(u_4-t)X(u_3)X(u_4)e^{-iwu_3}e^{iwu_4} \right)$$

$$\stackrel{(a)}{=} S_1 + S_2, \tag{21}$$

where (a) follows from the Isserlis' theorem [50] and $u_i \in [t - (N-1)/2 : t + (N-1)/2]$ for $i \in [1:4]$ with,

$$S_1 = \sum_{u_1,u_2,u_3,u_4} g_k(u_1 - t)g_k(u_2 - t)g_l(u_3 - t)g_l(u_4 - t)\mathsf{E}[X(u_1)X(u_3)]\mathsf{E}[X(u_2)X(u_4)]e^{-iwu_1}e^{iwu_2}e^{-iwu_3}e^{iwu_4},$$

$$S_2 = \sum_{u_1,u_2,u_3,u_4} g_k(u_1 - t)g_k(u_2 - t)g_l(u_3 - t)g_l(u_4 - t)\mathsf{E}[X(u_1)X(u_4)]\mathsf{E}[X(u_2)X(u_3)]e^{-iwu_1}e^{iwu_2}e^{-iwu_3}e^{iwu_4}.$$

$$S_1 = \sum_{v_1,v_2,v_3,v_4} g_k(v_1)g_k(v_2)g_l(v_3)g_l(v_4)e^{-iw(v_1+t)}e^{iw(v_2+t)}e^{-iw(v_3+t)}e^{iw(v_4+t)}$$

$$\cdot \int_{-\pi}^{\pi} A_{v_1+t}(\lambda)A_{v_3+t}^*(\lambda)e^{i\lambda(v_1+t)}e^{-i\lambda(v_3+t)}d\mu(\lambda) \int_{-\pi}^{\pi} A_{v_2+t}(\xi)A_{v_4+t}^*(\xi)e^{i\xi(v_2+t)}e^{-i\xi(v_4+t)}d\mu(\xi)$$

$$= \int_{-\pi}^{\pi} \sum_{v_1,v_3} g_k(v_1)g_l(v_3)e^{-iw(v_1+t)}e^{-iw(v_3+t)}A_{v_1+t}(\lambda)A_{v_3+t}^*(\lambda)e^{i\lambda(v_1+t)}e^{-i\lambda(v_3+t)}d\mu(\lambda)$$

$$\int_{-\pi}^{\pi} \sum_{v_2,v_4} g_k(v_2)g_l(v_4)e^{iw(v_2+t)}e^{iw(v_4+t)}A_{v_2+t}(\xi)A_{v_4+t}^*(\xi)e^{i\xi(v_2+t)}e^{-i\xi(v_4+t)}d\mu(\xi)$$

$$\stackrel{(a)}{=} \int_{-\pi}^{\pi} |A_t(\lambda)|^2 \sum_{v_1,v_3} g_k(v_1)g_l(v_3)e^{-iw(v_1+t)}e^{-iw(v_3+t)}e^{i\lambda(v_1+t)}e^{-i\lambda(v_3+t)}d\mu(\lambda)$$

$$\int_{-\pi}^{\pi} |A_t(\xi)|^2 \sum_{v_2,v_4} g_k(v_2)g_l(v_4)e^{iw(v_2+t)}e^{iw(v_4+t)}e^{i\xi(v_2+t)}e^{-i\xi(v_4+t)}d\mu(\xi) + O(B_{g_k}/B_X) + O(B_{g_l}/B_X)$$

$$= \int_{-\pi}^{\pi}\int_{-\pi}^{\pi} G_k(w-\lambda)G_l(w+\lambda)G_k(-w-\xi)G_l(-w+\xi)f_t(\lambda)d\lambda f_t(\xi)d\xi + O(B_{g_k}/B_X) + O(B_{g_l}/B_X), \tag{22}$$

where $(a)$ follows from Lemma 1. Similarly we have,

$$S_2 = \sum_{v_1,v_2,v_3,v_4} g_k(v_1)g_k(v_2)g_l(v_3)g_l(v_4) e^{-iw(v_1+t)} e^{iw(v_2+t)} e^{-iw(v_3+t)} e^{iw(v_4+t)}$$

$$\cdot \int_{-\pi}^{\pi} A_{v_1+t}(\lambda) A^*_{v_4+t}(\lambda) e^{i\lambda(v_1+t)} e^{-i\lambda(v_4+t)} d\mu(\lambda) \int_{-\pi}^{\pi} A_{v_2+t}(\xi) A^*_{v_3+t}(\xi) e^{i\xi(v_2+t)} e^{-i\xi(v_3+t)} d\mu(\xi)$$

$$= \int_{-\pi}^{\pi} \int_{-\pi}^{\pi} G_k(w-\lambda) G_l(-w+\lambda) G_k(-w-\xi) G_l(w+\xi) f_t(\lambda) d\lambda f_t(\xi) d\xi + O(B_{g_k}/B_X) + O(B_{g_l}/B_X). \quad (23)$$

$$\sum_{k,l=0}^{K-1} S_1 = \int_{-\pi}^{\pi} \int_{-\pi}^{\pi} \sum_{k=0}^{K-1} G_k(w-\lambda) G_k(-w-\xi) \sum_{l=0}^{K-1} G_l(w+\lambda) G_l(-w+\xi) f_t(\lambda) d\lambda f_t(\xi) d\xi + \sum_{k=0}^{K-1} O(B_{g_k}/B_X)$$

$$\leq \left( \int_{-\pi}^{\pi} \int_{-\pi}^{\pi} \left| \sum_{k=0}^{K-1} G_k(w-\lambda) G_k(-w-\xi) \right|^2 f_t(\lambda) d\lambda f_t(\xi) d\xi \right.$$

$$\left. \int_{-\pi}^{\pi} \int_{-\pi}^{\pi} \left| \sum_{l=0}^{K-1} G_l(w+\lambda) G_l(-w+\xi) \right|^2 f_t(\lambda) d\lambda f_t(\xi) d\xi \right)^{1/2} + \sum_{k=0}^{K-1} O(B_{g_k}/B_X). \quad (24)$$

where $(a)$ follows from (24) and a similar argument for $S_2$, as well as the orthogonality condition (15). This finishes the proof. ∎

**Yu Xiang** (S'10–M'15) received the B.S. degree with highest distinction in telecommunication engineering from Xidian University, Xi'an, China, in 2008, and the M.S. and Ph.D. degrees both in electrical and computer engineering from University of California, San Diego, CA, USA, in 2010 and 2015, respectively. From 2015 to 2018, he was a Postdoctoral Fellow with Harvard University, School of Engineering and Applied Sciences, Cambridge, MA, USA. He is currently an Assistant Professor with the Electrical and Computer Engineering Department, University of Utah, Salt Lake City. His research interests include statistical signal processing and information theory.

**Jie Ding** (M'12) received the Ph.D. degree in engineering sciences from Harvard University, Cambridge, MA, USA, in 2017, where he was also a Postdoctoral Researcher from January 2017 to December 2018. He was a Postdoctoral Researcher with the Rhodes Information Initiative, Duke University, Durham, North Carolina, USA, from January to August 2018. He is currently an Assistant Professor with the School of Statistics, University of Minnesota, Minneapolis, MN, USA. His research topics include signal processing, statistical inference, time series prediction, and machine learning with various applications.

**Vahid Tarokh** (F'09) received the Ph.D. degree in electrical engineering from the University of Waterloo, Waterloo, ON, Canada, in 1995. He was with AT&T Labs-Research and AT&T Wireless Services until August 2000 as Member, Principal Member of Technical Staff and, finally, as the Head of the Department of Wireless Communications and Signal Processing. In September 2000, he joined the Massachusetts Institute of Technology as an Associate Professor of Electrical Engineering and Computer Science. In June 2002, he joined Harvard University as a Gordon McKay Professor of Electrical Engineering and Hammond Vinton Hayes Senior Research Fellow. He was named Perkins Professor of Applied Mathematics in 2005. In January 2018, he joined Duke University, as the Rhodes Family Professor of Electrical and Computer Engineering, Computer Science, and Mathematics. From January 2018 to May 2018, he was also a Gordon Moore Distinguished Scholar in the California Institute of Technology.